\documentclass[12pt]{article}

\usepackage{hyperref}

\def\be{\begin{equation}}
\def\ee{\end{equation}}
\def\ba{\begin{array}}
\def\ea{\end{array}}
\def\d{\partial}
\def\dps{\displaystyle}

\begin{document}

\begin{flushright}
FIAN/TD/12/03
\end{flushright}

\vspace{1cm}

\begin{center}

{\bf \Large Two-column higher spin massless fields in $AdS_d$}

\vspace{.7cm}

\textsc{K.B. Alkalaev}

\vspace{.7cm}

{\em I.E. Tamm Department of Theoretical Physics, P.N. Lebedev Physical Institute,\\
Leninsky prospect 53, 119991, Moscow, Russia}

\vspace{3mm}
{\tt e-mail: alkalaev@lpi.ru}

\vspace{1cm}

\begin{abstract}

Particular class of $AdS_d$ mixed-symmetry bosonic massless
fields corresponding  to arbitrary two-column Young tableaux
is considered. Unique gauge invariant free actions are
found and equations of motion are analyzed.

\end{abstract}

\end{center}

\vspace{1.5cm}

\section{Introduction}

In the present paper we consider the problem of manifestly
covariant Lagrangian formulation of free mixed-symmetry
massless fields propagating on $d$-dimens\-ional anti-de
Sitter background. In contrast to flat spacetime, where
free higher spin dynamics is presently well understood
\cite{curt,AKO,Labastida,Siegel:1986zi,Pashnev,bekaert,hull},
the analysis of field dynamics on the $AdS_d$ background is
more complicated and up to now only particular examples of mixed-symmetry
fields were explicitly
analyzed \cite{BMV,Metsaev_d5,siegel,Zinoviev,Zinoviev2,ASV}.
The reason is that the classification of massless fields in
$AdS_d$ is essentially different from that for massless
fields in Minkowski space \cite{BMV}. From the
field-theoretical perspective this implies drastic changes
for the entire fabric of gauge symmetries as compared to
the field dynamics on the flat background
\cite{Metsaev,BMV}.

The line of consideration of the present paper is motivated
by the approach recently proposed in Ref. \cite{ASV}, which
represents mixed-symmetry  fields as gauge $p$-form
fields taking values in an appropriate
finite-dimensional irreps of $AdS_d$ algebra. The method was illustrated
by various examples of mixed-symmetry fields with
at most two rows. The goal of the present paper
is to apply the general prescription of
\cite{ASV} to higher spin massless fields
corresponding to  arbitrary two-column Young tableaux.

\section{Preliminaries}

Consider\footnote{Throughout the paper we work within the mostly minus signature
and use notations $\underline{m},\underline{n} = 0\div d-1\;$ for
world indices,
$a,b= 0\div d-1$ for tangent $so(d-1,1)$ vector indices and
$A,B = 0 \div d $ for tangent $so(d-1,2)$ vector indices.
We also use condensed notations for a set of antisymmetric indices:
$a [k]\equiv [a_1 \ldots a_k]$.
Indices denoted by the same letter are assumed to be antisymmetrized as
$X^a Y^a \equiv \frac{1}{2!}(X^{a_1}Y^{a_2}- X^{a_2}Y^{a_1}$).}
a Lorentz tensor field $\phi^{(s, p)}(x)$ on $d$-dimensional
spacetime (Minkowski or $AdS_d$)
\be
\label{field}
\phi^{a[s],\,b[p]}(x)\equiv \phi^{[a_1 \ldots a_s],\,[b_1 \ldots b_p]}(x)\;,
\ee
which is antisymmetric in both groups of indices, satisfies
the Young symmetry condition $\phi^{[\,a_1 \ldots a_s,\,a_{s+1}\,]b_2 \ldots
b_p}(x) = 0$, and contains all its traces.
The tensor $\phi^{(s,p)}(x)$ will be referred
to as the {\it metric-type} field \cite{ASV}.

The heights of columns $s$ and $p$ are assumed to satisfy
the following inequality
\be
\label{ineq}
0<p \leq s \leq \nu\;.
\ee
A value of the upper bound in (\ref{ineq}) is different
for massless fields in Minkowski and $AdS_d$. For the Minkowski space
$\nu\equiv\nu_{Mink}=\Big[\frac{d-2}{2}\Big]$ is a rank of the little Wigner group
$SO(d-2)$. For the $AdS_d$
space $\nu\equiv\nu_{AdS}=\Big[\frac{d-1}{2}\Big]$ is a rank of the
vacuum group $SO(d-1)$.
It follows that $\nu_{Mink}\leq \nu_{AdS}$, which allows one to conclude that theories
of massless fields in $AdS_d$ may give dual descriptions
of flat massless fields in the flat limit\footnote{For discussion of
dual formulations of free higher spin fields in Minkowski space
see papers \cite{hull2,hull1,bekaert,henneaux}.}.
If $s>\nu_{AdS}$ and $0< p\leq s$, the corresponding $AdS_d$ higher
spin theory is a dual formulation of some
particular $AdS_d$ theory with parameters $\tilde{p},\,\tilde{s}$ from (\ref{ineq}).

The metric-type field $\phi^{(s, p)}(x)$ (\ref{field})
is  a gauge field with the gauge transformation law
given by two types of gauge parameters \cite{Labastida} (schematically):
\be
\label{delta}
\delta\phi^{(s, p)}  = \d S^{(s-1,p)}+\d \Lambda^{(s, p-1)}  \;,
\ee
where tensors $S^{(s-1,p)}$ and $\Lambda^{(s, p-1)}$ are described by Young tableaux
obtained by cutting off a cell from the first and the second column of
$\phi^{(s, p)}(x)$, respectively, and contain all their traces.

In the sequel, we develop the  gauge invariant $AdS_d$
theory for a metric-type field $\phi^{(s, p)}(x)$  with arbitrary values of  $s$ and $p$ which satisfy (\ref{ineq}).
Before plunging into description of our
approach, discuss how one may proceed to obtain an $AdS_d$ theory
starting form some flat Lagrangian ${\cal L}$ invariant under
(\ref{delta}).
A general prescription consists in replacing $\d \rightarrow {\cal D}$, where
${\cal D}$ is the background Lorentz derivative commuting as
$[{\cal D},{\cal D}]\sim \lambda^2$, and adding appropriate mass-like
terms in ${\cal L}$ which support the gauge invariance. However, as pointed out in
\cite{Metsaev,BMV}, it is not always consistent procedure
for a generic mixed-symmetry field: only a part of
gauge symmetries can be deformed to $AdS_d$
to obtain a unitary theory. In the case under consideration, the symmetry which survives
in $AdS_d$ corresponds to the gauge parameter $\Lambda^{(s, p-1)}$. Lack of one of gauge symmetries on $AdS_d$
results in a discrepancy between degrees of freedom of a field $\phi^{(s, p)}(x)$
on the flat and $AdS_d$ backgrounds.
The balance may be restored by introducing a Stueckelberg
field for the missing symmetry $S^{(s-1,p)}$ which can be gauged away for $\lambda\neq 0$, however.
The flat limit of such extended theory describes not one but
two independent fields \cite{BMV, Zinoviev}.

Within our approach we do not  search for the $AdS_d$ deformation of some Lagrangian describing
flat field dynamics. Instead, we start with dynamics on the
$AdS_d$ background and investigate its flat limit then.

The background Minkowski or $AdS_d$ geometry is described by
the frame field $h^a= h_{\underline{n}}{}^a\,dx^{\underline{n}}$ and
Lorentz spin connection
$\omega^{ab} = \omega_{\underline{n}}{}^{ab}\,dx^{\underline{n}}$
which obey the equation
\be
\label{backgrounds}
[{\cal D}_{\underline{m}},{\cal D}_{\underline{n}}]\phi^{a[s],\, b[p]}
= \lambda^2 (s\,\,h_{\underline{m}}{}^{a}\, h_{\underline{n}\,c} \,\phi^{ca[s-1],\, b[p]}+p\,\,h_{\underline{m}}{}^{b}\, h_{\underline{n}\,c} \,\phi^{a[s],\, cb[p-1]} )
- (\underline{m}\leftrightarrow\underline{n})\;,
\ee
where
\be
\label{Lorentzderiv}
{\cal D}_{\underline{n}}\phi^{a[s],\, b[p]}
= \d_{\underline{n}}\phi^{a[s],\, b[p]}
+s\,\,\omega_{\underline{n}}{}^{a}{}_c\,\phi^{ca[s-1],\, b[p]}
+ p\,\,\omega_{\underline{n}}{}^{b}{}_c\,\phi^{a[s],\, cb[p-1]} \;,
\quad \d_{\underline{n}}=\frac{\d}{\d x^{\underline{n}}}\;.
\ee
Also, the zero-torsion condition
${\cal D}_{\underline{n}} h_{\underline{m}}{}^a
-{\cal D}_{\underline{m}} h_{\underline{n}}{}^a = 0$
is imposed. It expresses the spin connection
$\omega_{\underline{m}}{}^{ab}$ in terms of the first derivatives
of the frame field $h_{\underline{m}}{}^a$.
The equation (\ref{backgrounds}) describes $AdS_d$
spacetime with the symmetry algebra $o(d-1,2)$ when $\lambda^2>0$.
Minkowski space-time corresponds to $\lambda=0$.

The covariant D'Alembertian is
\be
\label{dal}
{\cal D}^2 \equiv {\cal D}^a{\cal D}_a =
h_{\underline{m}}^a {\cal D}^{\underline{m}}
(h_{\underline{n},\,a}{\cal D}^{\underline{n}})\;,
\ee
where the background Lorentz covariant derivative is given
by (\ref{Lorentzderiv}).

In what follows  we use
$AdS_d$ covariant notations and operate with $AdS_d$ tensors
 $T^{A[m]}(x)$. To relate Lorentz and $AdS_d$ covariant
realizations we introduce a compensator vector $V^A(x)$ normalized as
$V^A V_A = 1$ \cite{compensator}. This allows one to identify the Lorentz
subalgebra $so(d-1,1)$ within the $AdS$ algebra $so(d-1,2)$
as the stability algebra of the compensator, which results in the covariant
splitting of the $so(d-1,2)$ 1-form connection $\Omega^{[AB]}$
into the frame field $E^A$ and the Lorentz connection
$\omega^{[AB]}$: $E^A = DV^A\equiv dV^A+\Omega^{AB}V_B,\;$
$\omega^{[AB]} = \Omega^{[AB]} -2\lambda\,E^{[A}\,V^{B]}$
\cite{compensator}. In these notations, the background
$AdS_d$ geometry ($h^A$, $\omega_0^{[AB]}$) is defined by
the "zero-curvature" condition \cite{V_obz2}
\be
\label{zerocurv}
R^{AB}(\Omega_0) \equiv
d\Omega_0^{AB}+\Omega_0{}^A{}_C\wedge \Omega_0{}_{}^{CB}=0\;.
\ee
The action of the background covariant
derivative on an arbitrary $AdS_d$ tensor is given by \be
\label{backderaction} D_0 T^{A[m]} = d T^{A[m]} + m
\,\Omega_0^{A}{}_C\wedge T^{CA[m-1]}\;. \ee

\section{$p$-form gauge fields}

According to the general prescription of Ref. \cite{ASV}, we
describe two-column mixed-symmetry field
$\phi^{(s, p)}$ (\ref{field}) propagating on the $AdS_d$ space
in terms of a {\it frame-type} $p$-form gauge field
\be
\label{omega}
e^{a[s]}_{(p)} = e^{[a_1 ... a_s];\, [\underline {m}_1 ... \underline {m}_p ]}\;
dx_{\underline {m}_1} \wedge ... \wedge dx_{\underline {m}_p}\;.
\ee
It is convenient to replace world indices in (\ref{omega}) with
tangent indices
\be
e^{[a_1 ... a_s];\, [b_1 ... b_p ]}\equiv
e^{[a_1 ... a_s];\, [\underline {m}_1 ... \underline {m}_p ]}\;h_{\underline{m}_1}{}^{b_1} \ldots
h_{\underline{m}_p}{}^{b_p}\;,
\ee
where $h_{\underline{m}}{}^{a}$ is the background frame field in $AdS_d$.

The frame-type $p$-form gauge field (\ref{omega})
gives rise to a collection of components arising through
tensoring of world and tangent indices
\be
\label{dec}
e^{a[s];\,b[p]}
\sim \bigoplus_{i=0}^{p} \phi ^{a[s+i],\, b[p-i]}\;.
\ee
Here each tensor component $\phi ^{a[s+i],\, b[p-i]}$
corresponds to a Young tableau
with $(s+i)$ antisymmetric indices in the first column and $(p-i)$
antisymmetric indices in the second column and
contain all its traces. It is convenient to
denote each component in (\ref{dec}) as $\phi^{(i)}$. The first component $i=0$ in (\ref{dec}) is identified
with the metric-type field $\phi^{(0)}\equiv \phi^{(s, p)}$ (\ref{field}).

In principle, the $p$-form field (\ref{omega}) is sufficient to construct
a gauge invariant action functional. However, to control
gauge symmetries in a manifest manner one should introduce
additional $p$-form gauge fields.  In the case under
consideration, appropriate set of fields is given by
$e^{a[s]}_{(p)}$ and $\omega^{a[s+1]}_{(p)}$, which
will be referred to as the physical and the auxiliary
$p$-forms \cite{ASV}.

The Abelian curvature ($p+1$)-forms associated with the physical and
the auxiliary $p$-form gauge
fields read as
\be
\label{lor-curv}
r_{(p+1)}^{a[s]} = {\cal D} e^{a[s]}_{(p)} + h_b \wedge \omega^{a[s]b}_{(p)}\;,
\qquad
{\cal R}_{(p+1)}^{a[s+1]} = {\cal D} \omega^{a[s+1]}_{(p)} -(s+1)\lambda^2 h^a \wedge e^{a[s]}_{(p)}\;.
\ee
They are invariant under the gauge transformations with $(p-1)$-form gauge parameters
$\Lambda^{a[s]}_{(p-1)}$ and $\xi^{a[s+1]}_{(p-1)}$
\be
\label{gauge-lor}
\delta e_{(p)}^{a[s]} = {\cal D} \Lambda^{a[s]}_{(p-1)} + h_b \wedge \xi^{a[s]b}_{(p-1)}\;,
\qquad
\delta\omega_{(p)}^{a[s+1]} = {\cal D} \xi^{a[s+1]}_{(p-1)} -(s+1)\lambda^2 h^a \wedge \Lambda^{a[s]}_{(p-1)}\;.
\ee
In particular, the structure of the gauge invariance
(\ref{gauge-lor}) requires the curvatures to satisfy
Bianchi identities
\be
\label{bianchi}
{\cal D} r^{a[s]}_{(p+1)} + h_b \wedge {\cal R}_{(p+1)}^{a[s]b} =0 \;,
\qquad
{\cal D} {\cal R}_{(p+1)}^{a[s+1]} -(s+1)\lambda^2 h^a \wedge r^{a[s]}_{(p+1)} = 0\;.
\ee
The gauge transformations (\ref{gauge-lor}) are reducible. There is a set
of $(l+2)$-th level ($0\leq l \leq p-2$) gauge transformations of the form
\be
\label{reducible}
\ba{l}
\delta\Lambda_{(p-l-1)}^{a[s]} = {\cal D} \Lambda^{a[s]}_{(p-l-2)} + h_b \wedge \xi^{a[s]b}_{(p-l-2)}\;,
\\
\\
\delta\xi_{(p-l-1)}^{a[s+1]} = {\cal D} \xi^{a[s+1]}_{(p-l-2)} -(s+1)\lambda^2 h^a \wedge \Lambda^{a[s]}_{(p-l-2)}\;.
\ea
\ee
The role of the shift parameter $\xi^{a[s+1]}_{(p-1)}$ in the gauge transformations
(\ref{gauge-lor}) is to compensate all components of the physical $p$-form field
in (\ref{dec}) with $i>0$.
It can be easily seen from the decomposition of the gauge parameters $\xi^{a[s+1]}_{(p-1)}$
analogous to (\ref{dec})
\be
\label{gp2}
\xi^{a[s+1];\, b[p-1]}
\sim \bigoplus_{i=0}^{p-1} \xi ^{a[s+i+1],\, b[p-i-1]}\;,
\ee
where tensors in r.h.s. have Young symmetry properties
and contain all their traces.
Thus, by gauge fixing with the help of
the shift parameters $\xi^{a[s+1]}_{(p-1)}$,
the $p$-form gauge field (\ref{omega}) reduces
to the component $\phi^{(0)}$ corresponding to $i=0$ in (\ref{dec})
to be identified with the physical metric-type field (\ref{field}).
The derivative part of (\ref{gauge-lor}) can be  analyzed along the same lines. Namely,
introduce the decomposition of the derivative gauge
parameters  $\Lambda^{a[s]}_{(p-1)}$ analogous to (\ref{gp2})
\be
\label{gp1}
\Lambda^{a[s];\, b[p-1]} \sim \bigoplus_{i=0}^{p-1} \Lambda ^{a[s+i],\, b[p-i-1]}\;,
\ee
where tensors in r.h.s. have Young symmetry properties and
contain all their traces.
Then, one finds from
(\ref{dec}) and (\ref{gp1}) that
the metric-type field $\phi^{(0)} \equiv \phi ^{(s,p)}$
transforms as (schematically)
\be
\label{gau}
\delta \phi^{(0)}= {\cal D} \Lambda^{(0)}\;,
\ee
where the gauge parameter $\Lambda^{(0)}$ is the
first component in (\ref{gp1}) with $i=0$ and has the same
symmetry type as $\Lambda^{(s,p-1)}$ in (\ref{delta}). As a consequence of the
(\ref{reducible}), the gauge transformation (\ref{gau}) is
reducible up to a $p$-th level.

It is worth to comment that our
approach is formulated in the way it incorporates  the
gauge symmetry with parameter $\Lambda^{(s,p-1)}$
(\ref{delta}) only, which is the correct gauge symmetry
on the $AdS_d$ background. Another type of gauge symmetry  with parameter
$S^{(s,p-1)}$ (\ref{delta}) is not placed in the $AdS_d$
formulation and appears in the flat limit $\lambda=0$.
We shall comment on this phenomenon later.

The Lorentz $p$-form gauge fields introduced to describe a mixed-symmetry
field in $AdS_d$ can be viewed as a result of the decomposition
with respect to the Lorentz group of a $p$-form gauge field carrying an appropriate irreducible representation of $o(d-1,2)$
\cite{ASV}. In our case,
the Lorentz fields $e_{(p)}^{a[s]}$ and $\omega_{(p)}^{a[s+1]}$ result
form the following $AdS_d$ $p$-form field
\be
\label{adslor}
\Omega_{(p)}^{A[s+1]} \sim e^{a[s]}_{(p)} \oplus \omega^{a[s+1]}_{(p)} \;.
\ee
With the help of the compensator vector discussed in section 2, the isomorphism (\ref{adslor}) takes the precise form
\be
\label{adslorcor}
\Omega_{(p)}^{A[s+1]} = \omega^{A[s+1]}_{(p)} +\lambda\,(s+1)\, V^A\, e^{A[s]}_{(p)} \;,
\ee
supplemented with the transversality conditions
\be
e^{A[s-1]C}_{(p)}V_C = 0 \;,\quad \omega^{A[s]C}_{(p)}V_C =0 \;.
\ee
The appearance of the cosmological parameter $\lambda$ in (\ref{adslorcor}) is motivated by different
mass dimensions of the physical and the auxiliary $p$-forms since
on the level of equations of
motion the auxiliary field is expressed in terms of the
first derivatives of the physical field.

In the $AdS_d$ formalism, the gauge transformation and the curvature are given by
($0\leq l \leq p-2$)
\be
\label{curv}
R_{(p+1)}^{A[s+1]} = D_0 \Omega_{(p)}^{A[s+1]}\;,
\qquad \delta \Omega_{(p)}^{A[s+1]} = D_0 \xi_{(p-1)}^{A[s+1]}\;,
\quad
\delta\xi_{(p-l-1)}^{A[s+1]} = D_0 \xi_{(p-l-2)}^{A[s+1]}\;,
\ee
where
\be
R_{(p+1)}^{A[s+1]} \sim r_{(p+1)}^{a[s]}\oplus {\cal R}_{(p+1)}^{a[s+1]}\;,
\qquad
\xi_{(p-l-1)}^{A[s+1]} \sim \Lambda_{(p-l-1)}^{a[s]}\oplus \xi_{(p-l-1)}^{a[s+1]}\;
\ee
and the covariant derivative $D_0$ is evaluated
with respect to the background $AdS_d$ connection $\Omega_0$ (\ref{backderaction}).
Bianchi identities
\be
D_0 R_{(p+1)}^{A[s+1]} = 0\;
\ee
are the consequence of the zero-curvature condition $D_0^2=0$ (\ref{zerocurv}).

It is interesting to note  that in
$d=7 \;mod\; 4$ dimensions one more irreducibility condition  on the tangent
indices may be imposed. A duality  condition may arise
if the tangent $so(d-1,2)$ Young tableau is a column of maximal height
$s+1=\frac{d-1}{2}+1$, which may be dualized using Levi-Civita symbol
\be
\label{duality1}
\Omega_{(p)}{}_{A[s+1]}= \pm\frac{1}{(s+1)!}\;
\epsilon_{A[s+1]B[s+1]}\; \Omega_{(p)}^{B[s+1]}
\;.
\ee
This condition decomposes $\Omega_{(p)}^{A[s+1]}$ into selfdual and antiselfdual parts and
results in the following relation for Lorentz $p$-form fields
\be
\label{duality2}
\omega_{(p)}{}^{a[s+1]} = \pm\frac{\lambda}{s!}\;
\epsilon^{a[s+1]b[s]}\,e_{(p),\;b[s]}\;.
\ee
This off-shell  constraint can be used to  express algebraically
the auxiliary field in terms of the physical field or {\it vice versa}.
In this case equations of motion will acquire first order
form. Note that in $d=5 \;mod\; 4$ dimensions the duality
condition may be also imposed but then
the field $\Omega_{(p)}^{A[s+1]}$ is necessarily complex.
The analogous phenomenon
was discussed for massive antisymmetric tensor fields
on the flat background  in \cite{dual}
and called odd-dimensional selfduality. In the $AdS_d$ theory
the role of mass is played by the cosmological constant $\lambda$.

\section{Dynamics}

In what follows we build higher spin action
which correctly describes free dynamics of metric-type field $\phi^{(s, p)}$ on
the $AdS_d$ background. It was suggested in \cite{ASV} to look for
free action in the MacDowell-Mansouri form \cite{LV,MacDowell:1977jt}.
In our case the most general action reads
\be
\label{action}
\ba{c}
\dps
{\cal S}_{2} =
\frac{\kappa_1}{\lambda^2}\int_{{\cal M}_d} H_{A[2p+2]}\,
\wedge\;
R_{(p+1)}^{A[p+1]B[s-p]}
\wedge\;
R_{(p+1)}^{A[p+1]}{}_{B[s-p]}
\\
\\
\dps
+
\frac{\kappa_2}{\lambda^2} \int_{{\cal M}_d} H_{A[2p+2]}\,
\wedge\;
R_{(p+1)}^{A[p+1]B[s-p-1]C}
\wedge\;
R_{(p+1)}^{A[p+1]}{}_{B[s-p-1]}{}^{D}V_C V_D\;,

\ea \ee
where $\kappa_{1,2}$ are arbitrary dimensionless constants and the notation
is introduced:
\be
H_{A[m]} =\epsilon_{A_1 \ldots A_m B_{m+1}  \ldots B_{d+1}}
h^{B_{m+1}}\wedge\ldots\wedge h^{B_d}V^{B_{d+1}}\,.
\ee
The freedom in $\kappa_{1,2}$  can be fixed up to an overall factor in front of the action
(\ref{action}) by adding a unique total derivative term
\be
\ba{c}
\dps
{\cal O} = \int_{{\cal M}_d} d\,\Big(\;H_{A[2p+3]}\,
\wedge\;
R_{(p+1)}^{A[p+2]B[s-p-1]}
\wedge\;
R_{(p+1)}^{A[p+1]}{}_{B[s-p-1]}{}^C V_C\,\Big)\;.
\ea \ee

Note that the form of the action imposes a
natural restriction $p\leq [\frac{d-2}{2}]$, while $s$ is not
restricted at all. It is tempting to speculate that for
$s>\nu_{AdS}$, the action (\ref{action})
gives rise to dual formulations of two-column metric-type
fields in $AdS_d$.

The variation of the action (\ref{action})
gives rise to the equations of motion
\be
\label{eom_ads}
\epsilon^{A[p+1]}{}_{B[p+1]C[d-2p-1]}h^{C_1}\wedge\ldots\wedge h^{C_{d-2p-1}}
\wedge R_{(p+1)}^{A[s-p]B[p+1]} = 0\;.
\ee
To clarify the dynamical content hidden in these equations it is convenient to
perform analysis in terms of Lorentz components. To this end, introduce
the decomposition of the curvature tensors (\ref{lor-curv}) analogous to (\ref{dec})
\be
\label{ct1}
r^{a[s];\, b[p+1]} \sim \bigoplus_{i=0}^{p+1} r^{a[s+i],\, b[p-i+1]}\;,
\ee
\be
\label{ct2}
{\cal R}^{a[s+1];\, b[p+1]} \sim \bigoplus_{i=0}^{p+1} {\cal R}^{a[s+i+1],\, b[p-i+1]}\;,
\ee
where tensors in r.h.s.-s  have Young symmetry properties
and contain all their traces.
By direct calculation one shows that equations  (\ref{eom_ads}) can be
cast into the form
\be
\label{Weyl1}
\ba{l}
p<s: \quad r^{a[s]}_{(p+1)} = h_{b_1}\wedge \ldots \wedge h_{b_{p+1}}\,T^{a[s],\,b[p+1]}\;,\qquad
\\
\\
p=s:\quad  r^{a[s]}_{(p+1)} = 0\;
\ea
\ee
and
\be
\label{Weyl2}
p\leq s:\quad {\cal R}^{a[s+1]}_{(p+1)} = h_{b_1}\wedge \ldots \wedge h_{b_{p+1}}\,C^{a[s+1],\,b[p+1]}\;
\ee
with arbitrary traceless 0-forms $T^{a[s],\,b[p+1]}$ and
$C^{a[s+1],\,b[p+1]}$ described by two-column Young tableaux.
All other components of the curvatures (\ref{ct1})-(\ref{ct2}) are zero.
One should note that the tracelessness for $T^{a[s],\,b[p+1]}$
is required only when $\lambda\neq 0$ (it follows from Bianchi
identities):
\be
\label{trT}
\lambda \,T^{a[s-1]c\,,\, b[p]d}\eta_{cd}=0\;.
\ee
We shall discuss  the distinguished role of this
condition later.

In the case $p=s$, which corresponds to the physical
metric-type field described by rectangular Young tableau,
0-form  $C$ is the \textit{primary} Weyl tensor
\cite{ASV}. It parameterizes those components of the
curvatures which are non-vanishing on the mass-shell and cannot
be expressed through derivatives of some other
curvature components.
In the case of the physical
metric-type field described by
non-rectangular Young tableau with $p<s$, the
\textit{primary} Weyl tensor is given by 0-form $T$, while the 0-form $C$
is the \textit{secondary} Weyl tensor expressed
through the first derivatives of the primary Weyl tensor $T$ by virtue of
Bianchi identities \cite{ASV}. The structure of Weyl tensors, both primary
and secondary, is in agreement with Young symmetry types of
invariant curvature tensors found in \cite{hull} in the framework of non-local
formulation of Minkowski higher spin dynamics.

Turn now to the explicit analysis of the equations (\ref{Weyl1})-(\ref{Weyl2}).
By analogy with the decomposition (\ref{dec}), introduce the decomposition
of the auxiliary $p$-form:
\be
\label{dec_aux}
\omega^{a[s+1];\, b[p]}
\sim \bigoplus_{i=0}^{p} \omega ^{a[s+i+1],\, b[p-i]}\;.
\ee
Then, one shows that all components
$\omega^{(i)}$ (\ref{dec_aux}) of the auxiliary $p$-form can be
expressed  in terms
of the first derivatives of the components $\phi^{(i)},\; 0\leq i \leq p$ of
the physical $p$-from (schematically):
\be
\label{gt_aux}
\omega^{(i)} = {\cal D}\phi^{(i)} + {\cal D}\phi^{(i+1)}\;,\qquad 0\leq i \leq p\;.
\ee
This expression follows from the fact that the auxiliary
field enters the physical curvature (\ref{lor-curv})
without derivatives and the equations (\ref{Weyl1}) are, in
fact, linear homogeneous equations with respect to the
components of the auxiliary $p$-form. Notice that when
$p<s$, the curvature component $T^{a[s],\,b[p+1]}$ is not
zero and does not contain any components of the auxiliary
$p$-form (\ref{dec_aux}). Therefore, the number of linear
homogeneous equations (\ref{Weyl1}) matches  exactly the
number of components of the auxiliary $p$-form. In other
words, Eq.(\ref{Weyl1}) is the
constraint, which allows one to express the auxiliary
field in terms of the first derivatives of the physical
field.

As discussed in section 3, the specific feature of
the gauge parameter $\xi_{(p-1)}^{a[s+1]}$ is that it
enters the gauge transformations (\ref{gauge-lor})
algebraically for the dynamical $p$-form and
through a derivative for the auxiliary $p$-form.
From (\ref{gp2}) it follows that the gauge transformations (\ref{gauge-lor})
can be cast into the schematic form
\be
\ba{l}
\delta \phi^{(i)} = \xi^{(i-1)}\;,\qquad 1\leq i \leq p\;,
\\
\\
\delta \omega^{(j)} = {\cal D}\xi^{(j)}+{\cal D}\xi^{(j-1)}\;,\qquad 0\leq j \leq p\;.
\ea
\ee
Combining these expressions with (\ref{gt_aux})
one finds that, by gauge fixing all redundant
components of the physical $p$-form field to zero,
the auxiliary field expresses through the physical one as
\be
\label{aux_ph}
\omega^{(0)} = {\cal D}\phi^{(0)}\;.
\ee

As a consequence of (\ref{Weyl2}), the second-order dynamical equations of motion on the physical metric-type
component $\phi^{(s,p)}$ emerge as the trace of the component ${\cal R}^{(0)}$ (\ref{ct2})
\be
\label{eq-tr}
{\cal R}^{a[s]c\,,}{}_c{}^{\,b[p]} =0\;,
\ee
where the auxiliary field is expressed through the first derivatives
of the physical one according to (\ref{aux_ph}).
Clearly, the tensor in l.h.s. of (\ref{eq-tr})
has the same Young symmetry type  as that of the metric-type field
$\phi^{(s,p)}$.

To find the explicit form of dynamical
equations of motion, we simplify our calculations
introducing Fock space notations for metric-type fields
\be
\label{mtfock}
|\phi\rangle = \phi_{[a_1 \ldots a_s],\,[b_1 \ldots b_p]}\; \alpha_1^{a_1}\ldots \alpha_1^{a_s} \alpha_2^{b_1}\ldots \alpha_2^{b_p}|0\rangle\;,
\quad
\ee
where $\alpha_i^a$ and $\bar{\alpha}_j^b$, $\;i,j=1,2$
are creation and annihilation operators
defined on the Fock vacuum
\be
\bar{\alpha}^i_c |0\rangle =0\;
\ee
and satisfying the  algebra
\be
\{\bar{\alpha}^i_a, \alpha^j_b  \} = \delta^{ij}\eta_{ab}\;,
\quad
\{\alpha^i_a, \alpha^j_b \} = 0\;,
\quad
\{\bar{\alpha}^i_a, \bar{\alpha}^j_b \} =0\;.
\ee
The following notations are convenient in practice \be
\label{f_operators} \ba{c} L_{ij}=
\bar{\alpha}_i^a\bar{\alpha}_j{}_a\;, \quad
T_{ij}=\alpha_i^a\bar{\alpha}_j{}_a\;, \qquad
N_{ij}=\alpha_i^a \alpha_j{}_a\;,
\\
\\
D_i = \alpha_i^a\,{\cal D}_a\;,
\quad
\bar{D}_i = \bar{\alpha}_i^a\,{\cal D}_a\;.
\\
\\
\ea
\ee
Vector $|\phi\rangle$ satisfies the conditions
\be
T_{12}|\phi\rangle =0\;,
\quad
T_{11}|\phi\rangle = s_1 |\phi\rangle\;,
\quad
T_{22}|\phi\rangle = s_2 |\phi\rangle\;,
\ee
which reflect its Young symmetry properties.

The physical metric-type field (\ref{field})
expressed through the $p$-form field (\ref{omega})
$|e\rangle = e_{[a_1 \ldots a_s];\,[b_1 \ldots b_p]}\; \alpha_1^{a_1}\ldots \alpha_1^{a_s} \alpha_2^{b_1}\ldots \alpha_2^{b_p}|0\rangle\;$
has the form
\be
|\phi\rangle =
\sum_{k=0}^{p}(-)^k\frac{(s-k)!}{k!}\,
\alpha_1^{a_1}\ldots \alpha_1^{a_k} \,
\alpha_2^{b_1}\ldots \alpha_2^{b_k}\,
\bar{\alpha}_1{}_{\,b_1} \ldots \bar{\alpha}_1{}_{\,b_k} \,
\bar{\alpha}_2{}_{\,a_1}\ldots  \bar{\alpha}_2{}_{\,a_k}\,
|e\rangle\;.
\ee
With the help of operators (\ref{f_operators})
the auxiliary field (\ref{aux_ph}) and
the gauge transformation (\ref{gau}) acquire
exact form
\be
\label{f-aux}
|\omega\rangle = D_1|\phi\rangle\;,
\ee
\be
\label{ads-sym}
\delta|\phi\rangle = \Big(D_2
-\frac{1}{s-p+1}\,D_1 T_{21}\Big)|\Lambda\rangle\;.
\ee
The $i=0$ component $|{\cal R}^{(0)}\rangle$ of the decomposition
(\ref{ct2}) is
\be
\label{R-aux}
|{\cal R}^{(0)}\rangle = \Big( D_2 - \frac{1}{s-p+1}\, D_1 T_{21} \Big)
|\omega\rangle-\lambda^2(s-p+2)N_{12}|\phi\rangle\;.
\ee
Substituting the expression for the auxiliary field (\ref{f-aux}) into
the curvature (\ref{R-aux}) and taking  trace
$L_{12}|{\cal R}^{(0)}\rangle = 0$ (\ref{eq-tr})
we find the equation of motion
\be
\label{equation}
\Big({\cal D}^2-D_1 \bar{D}_1 - D_2\bar{D}_2
-D_1 D_2 L_{12}
+\lambda^2 N_{12} L_{12}
+\lambda^2(d(p-1)+2p-p^2+s)\Big)|\phi\rangle= 0\;.
\ee
The covariant D'Alembertian ${\cal D}^2$ is given by (\ref{dal}).
In the gauge $\bar{D}_i|\phi\rangle=0$ and
$L_{12}|\phi\rangle = 0$ we are left with
\be
\label{alk}
\Big({\cal D}^2+\lambda^2(d(p-1)+2p-p^2+s)\Big)|\phi\rangle= 0\;.
\ee
To compare these equations with the previously
known results, we reproduce here the higher spin equations of
\cite{Metsaev} written in a covariant gauge
for  arbitrary mixed-symmetry massless fields
$\phi^{(h_1,\ldots,h_\nu)}$
corresponding to the $AdS_d$ representations $D(E_0, {\bf s})$ with
the energy $E_0$ saturating the unitary bound and
spin ${\bf s} = (h_1, \ldots, h_\nu)$ (here $h_l$
are lengths of rows of corresponding traceless $o(d-1)$ Young tableau and
$\nu\equiv \nu_{AdS}$ is a rank of $o(d-1)$)
\be
\label{mets}
\Big({\cal D}^2-\lambda^2(h_k-k-1)(h_k-k-2+d)+\lambda^2\sum_{l=1}^{\nu}h_l)\Big)\phi^{(h_1,\ldots,h_\nu)}= 0\;.
\ee
Here $h_k$ is the length of upper rectangular block and $k$ is the number
of the bottom row in this block ({\it i.e.}, $k$ is a height of the block). The case of two-column Young tableaux corresponds to
\be
\label{spin}
k=p\;,\,\quad h_k = 2\;,\qquad h_l= \left\{
\ba{l}
2\,,\quad 1 \leq l\leq p\,
\\
1\,,\quad p < l\leq s\,
\ea\;.
\right.
\ee
Plugging these values into (\ref{mets}) one indeed arrives
at the equations (\ref{alk}).

One should note that the equations (\ref{mets}) are
true in even dimensions for any $h_l$ and in odd
dimensions when $h_{(d-1)/2}=0$ \cite{Metsaev}, and
should not be taken for granted for selfdual
and antiselfdual representations which appear for non-zero
$\pm h_{(d-1)/2}$. In our formulation $h_{(d-1)/2}$ can
take non-zero values and, in fact, enters the theory only
through $|h_{(d-1)/2}|$. We suggest our equations describe
a sum of irreps with $-h_{(d-1)/2}$ and $+h_{(d-1)/2}$
and the duality condition (\ref{duality1}) discussed above
may help to extract selfdual and antiselfdual parts corresponding to
different signs of $h_{(d-1)/2}$.

Discuss now the flat limit $\lambda=0$ of the
field equations (\ref{equation}). It turns
out that new symmetry $\delta|\phi\rangle =\d_1  |S\rangle$ appears in addition to that defined by (\ref{ads-sym})
and the general gauge transformation takes now the form
\be
\label{flat-sym}
\delta|\phi\rangle =\d_1  |S\rangle +
\Big(\d_2
-\frac{1}{s-p+1}\,\d_1 \,T_{21}\Big)|\Lambda\rangle\;,
\ee
where Fock vector $|S\rangle\ $ is associated with the tensor
$S^{(s-1,p)}$ (\ref{delta})
and the operators $\d_i$ are obtained from $D_i$-s (\ref{f_operators}) by replacing
${\cal D}^a\rightarrow \d^a$. One can check directly
that the equations (\ref{equation}) are
do invariant under the gauge transformation (\ref{flat-sym}) with
the gauge parameter  $|S\rangle\ $ at $\lambda=0$.
However, the simplest way to prove the invariance
is to observe that the auxiliary field being expressed through the physical
field is an invariant expression with respect to
$\delta|\phi\rangle =\d_1  |S\rangle$. Consequently, this additional
invariance follows for the equations of motion on the flat background.

Note that starting from the equations of motion (\ref{equation})
one can develop an extended theory in the
$AdS_d$ space which realizes missing symmetry $|S\rangle\ $ as Stueckelberg symmetry \cite{BMV}.
To this end one should perform a shift
$|\phi\rangle  \rightarrow|\phi\rangle + D_1  |\chi\rangle $  inside the
equation of motion (\ref{equation}) with
$|\chi\rangle$ having Young symmetry type of $|S\rangle$.
Besides its own symmetries with derivatives, the Stueckelberg field
$|\chi\rangle$ will be transformed as
$\delta |\chi\rangle = \lambda|S\rangle$.
The resulting equation involving  $|\phi\rangle$ and $|\chi\rangle$ will be invariant
under restored gauge symmetry with parameters $|\Lambda\rangle$ and $|S\rangle$.
To obtain the  equation for the Stueckelberg field $|\chi\rangle$,
one should make the same shift inside the tensor
$\lambda \,tr \,| T \rangle =0$ (\ref{trT}).
As a result one will get the second-order equation for
the Stueckelberg field containing first-order terms with respect to
the field $|\phi\rangle$.
Then, the role of the condition (\ref{trT}) becomes
quite clear. This is the differential condition (Noether identity)
for a leftover symmetry with the parameter $|S\rangle$ (for more details see \cite{BMV}).
In the flat limit  equations become diagonal with respect to
fields $|\phi\rangle$ and $|\chi\rangle$ and describe a sum
of independent higher spin fields.

\section{Concluding remarks}

In this paper we described higher spin fields in $AdS_d$ corresponding to arbitrary
two-column Young tableaux. Our approach is a particular
realization of the general scheme proposed in
\cite{ASV} for generic mixed-symmetry fields. The
method employs the set of $p$-forms with antisymmetric
tangent indices having different dynamical interpretation.
So, one distinguishes between the physical
field and the auxiliary field and the latter is expressed
in terms of the physical field by virtue of its equations
of motion. The system has
the gauge symmetry relevant to the anti-de Sitter
background and reveals an additional symmetry in the flat
limit $\lambda = 0$.

To conclude, it is worth to comment that the flat limit of the action
functional (\ref{action}) may give rise to the
dual descriptions of higher spin fields in Minkowski space
\cite{hull2,hull1,bekaert,henneaux}. The particular
example is the three-cell ``hook'' field in $AdS_5$.
In the flat limit the massless ``hook'' field decomposes
into a couple of spin-2 massless fields because massless
fields in $5d$ Minkowski space are exhausted by totally
symmetric fields. As is easily seen from the equations
(\ref{equation}) (see also Ref. \cite{ASV}), these spin-2
massless fields are formulated in terms of dual
variables \cite{curt,hull2,henneaux}.
In particular, the curvature tensors (\ref{Weyl1}) and
(\ref{Weyl2}) become related by the duality transform
as in \cite{hull2}.

\vspace{1.5cm}

{\bf Acknowledgements}
\\
\\
I am grateful to O.V. Shaynkman and M.A. Vasiliev for
useful discussions. Also, I would like to thank O.V.
Shaynkman for careful reading of the manuscript and useful
remarks. This work was partially supported by RFBR Grants
No 02-02-16944, MAC No 03-02-06462, LSS-1578-2003-2 and the
Landau Scholarship Foundation, Forschungszentrum
J\"u\-lich.

\end{document}